\documentclass[aps,prl,twocolumn,showpacs,amsmath,amssymb,floatfix]{revtex4}
\usepackage{epsfig}
\usepackage{bm}

\begin{document}

\title{Shot noise thermometry of the quantum Hall edge states}

\author{Ivan P. Levkivskyi and Eugene V. Sukhorukov}
\affiliation{D\'epartement de Physique Th\'eorique, Universit\'e de Gen\`eve, CH-1211 Gen\`eve 4, Switzerland}
\date{\today}

\begin{abstract}
We use the non-equilibrium bosonization technique to investigate effects of the Coulomb interaction on  
quantum Hall edge states at filing factor $\nu=2$, partitioned by a quantum point contact (QPC). We find, 
that due to the integrability of charge dynamics, edge states evolve to a non-equilibrium stationary 
state with a number of specific features. In particular,
the noise temperature $\Theta$ of a weak backscattering current between edge channels is linear in voltage 
bias applied at the QPC, independently of the interaction strength. In addition, it is a non-analytical 
function of the QPC transparency $T$ and scales as $\Theta \propto T\ln(1/T)$ at $T\ll 1$. 
Our predictions are confirmed by exact numerical calculations.
\end{abstract}


\maketitle

\begin{figure}[h]
\begin{center}
\epsfxsize=7.5cm
\epsfbox{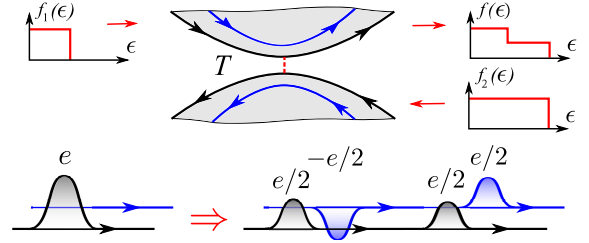}
\end{center}
\vspace{-3mm}
\caption{(Color on-line) Fermionic and bosonic aspects of the edge states physics.
{\it Upper panel:} At zero temperature, the electron distribution functions of the edge states arriving 
at the biased QPC are $f_1(\epsilon) = \theta(\epsilon_F-\epsilon)$ and 
$f_2(\epsilon) = \theta(\epsilon_F +\Delta\mu- \epsilon)$. If the transparency $T$ of the QPC is independent
of the energy, then the distribution function of the outgoing electrons is 
$f(\epsilon) = (1-T)f_1(\epsilon)+Tf_2(\epsilon)$. {\it Lower panel:} 
Schematic illustration of the strong Coulomb interaction effect at the QH edge at filling factor 
$\nu=2$. The electron wave packet of the charge $e$, created in the outer edge channel (lower, black line), 
decays into two eigenmodes of the edge Hamiltonian, the charged and dipole mode, which
propagate with different speeds and carry the charge $e/2$ in the outer channel.
Similar situation arises when an electron is injected in the 
inner channel (upper, blue line).} \vspace{-3mm}
\label{conc}
\end{figure}

Rapid experimental progress in the field of the electron transport in one-dimensional systems 
has unveiled new exciting phenomena inherent in strong, non-perturbative interactions 
characteristic of such systems. The notable examples are the recent experiments on the energy 
relaxation \cite{exper}, and on non-equilibrium dephasing of quantum Hall (QH) edge states 
\cite{firstMZ, MZ}. These chiral electron states 
may be viewed as quantum analogs of classical skipping orbits arising at the edge of a two dimensional 
electron system exposed to a perpendicular magnetic field. The aforementioned experiments utilize QPCs to 
bring edge states of opposite chirality close to each other in order to mix them, thereby inducing 
electron backscattering. By applying a voltage bias between these edge states, one may create  
a non-equilibrium state with the electron distribution function in the form of a 
``double-step'' \cite{exper} (see the upper panel of Fig.~\ref{conc}). 

The double-step distribution is characteristic of the effectively free-fermion behavior of 
electrons in metals \cite{explain}. Weak interactions leads to the equilibration of electrons 
in the long-time limit. At the QH edge, however, this distribution may evolve in a non-trivial 
way \cite{others} and, in the weak injection regime, through several intermediate asymptotics \cite{our-rel},
before reaching the equilibrium state. 
At the origin of this behavior are the non-perturbative interaction effects: For the Landau level 
filling factor $\nu>1$, when several co-propagating channels coexist at the edge, 
the strong Coulomb interaction leads to the formation of collective excitations called 
edge magneto-plasmons \cite{group2} (see the lower panel of Fig.~\ref{conc}). 
Propagating with different velocities, these excitations strongly redistribute electrons.
We have shown earlier \cite{our}, that this process is also responsible 
for non-monotonic dephasing observed in the resent experiments \cite{MZ}. 

\begin{figure}[h]
\begin{center}
\epsfxsize=8cm
\epsfbox{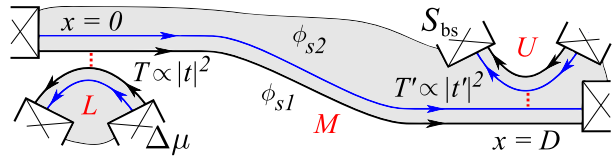}
\end{center}
\vspace{-3mm}
\caption{(Color on-line) Schematics of the measurement of the effective noise temperature. 
The ``double-step'' distribution is created at the left ($x = 0$) voltage-biased QPC of the arbitrary 
transparency $T$. The state propagates towards the right ($x=D$) QPC of the small transparency 
$T'\ll 1$, connected to a cold Ohmic contact, and induces the zero-frequency backscattering current noise, 
$S_{\rm bs}$. Thereby, the right QPC serves as a detector of the effective temperature of this noise, 
$\Theta\propto S_{\rm bs}$. 
The notations for the boson fields describing each QH edge are shown near the corresponding 
edge channels: the index $s = L,M,U$ enumerates the edges, while the index $\alpha = 1,2$ enumerates the edge 
channels at the same edge at filling factor $\nu =2$.} \vspace{-3mm}
\label{tun}
\end{figure}

Instead of  determining directly the electron distribution function, as in Ref.\ 
\cite{exper}, one may investigate the  effects of interactions in a non-equilibrium state 
by measuring the effective noise temperature of a system \cite{Blanter}. 
One way of doing this in a QH system \cite{Heiblum} is by attaching a cold Ohmic contact
to the co-propagating edge channel,
via the second QPC, as shown in Fig.\ \ref{tun}, and measuring 
the zero-frequency noise power $S_{\rm bs}$ of the backscattering current 
$j_{\rm bs}$: 
\begin{equation}
S_{\rm bs}=\int dt\langle j_{\rm bs}(t) j_{\rm bs}(0)\rangle.
\label{zfnp}
\end{equation}
The important property of this measurement scheme is that in the absence of the interaction between 
the channels, one should not expect any influence of the electron injection at the first, source, QPC
on the noise at the second, detector, QPC. Therefore, by measuring $S_{\rm bs}$ as a function of the 
voltage bias $\Delta\mu$, the transparency $T$ of the source QPC, and of the distance $D$ between 
the QPCs, one may investigate interaction effects and the evolution of a non-equilibrium state 
initially prepared at the first QPC.  In this Letter, we demonstrate that the strong interaction and 
the integrability of the charge dynamics at the QH edge lead to the formation of a non-equilibrium stationary 
state, which manifests itself in the singular, non-analytical behavior of the effective noise temperature.    

{\it Effective noise temperature.}---
In the regime of weak tunneling at the second, detector QPC, one can write \cite{Blanter}:   
\begin{equation}
S_{\rm bs} = G_D\int d\epsilon \{f(\epsilon)[1-f_D(\epsilon)] + f_D(\epsilon)[1-f(\epsilon)]\},
\label{useful}
\end{equation}
where $G_D$ is the conductance of the QPC, $f(\epsilon)$ is the electron distribution function 
in the inner channel, and $f_D(\epsilon)$ is the equilibrium distribution in the detector's Ohmic contact.
Assuming the Fermi distribution $f(\epsilon)=f_F(\epsilon-\epsilon_F)$ with the temperature 
$\Theta_{\rm eq}$ in the inner edge channel, and with the zero temperature at the detector's Ohmic contact,  
$f_D(\epsilon)=\theta(\epsilon_F-\epsilon)$, one immediately finds that $S_{\rm bs} = (2\ln2)G_D\Theta_{\rm eq}$.
Therefore, away from equilibrium, it is natural to define the effective noise temperature 
$\Theta$ via the expression 
\begin{equation}
S_{\rm bs}\equiv (2\ln 2 )G_D\Theta.
\label{definition}
\end{equation}
On the other hand, since the inner and the outer edge channels are electrically isolated
from each other, there is no average current contribution from the first QPC to the inner channel, 
which may be expressed
as $\int d\epsilon [f(\epsilon)-\theta(\epsilon_F-\epsilon)]=0$. Combining this identity with the 
expression (\ref{useful}), one obtains the simple expression for the effective noise temperature:
\begin{equation}
\Theta = (1\!/\!\ln 2)\int_{\epsilon_F}^\infty d\epsilon f(\epsilon).
\label{t2f}
\end{equation}

Facing strong interactions that cannot be accounted for perturbatively, one may choose 
to treat tunneling
at the first QPC perturbatively with respect to its small transparency $T$. Recently,
using this method, the Ref.\ \cite{neder-wrong} 
has found that the noise temperature $\Theta$ is linear in $T$, while non-perturbative interactions
manifest themselves in the non-trivial power-law dependence of $\Theta$ on the voltage bias $\Delta\mu$. 
However, it turns out that far from the injecting QPC, where a non-equilibrium stationary
state arises,  the perturbation theory fails to correctly describe the behavior of $\Theta$ at small $T$.
Very roughly, this happens because the weak partitioning noise at the first QPC generates a correction 
to the distribution function of the form
$f(\epsilon)\propto T\Delta\mu /(\epsilon-\epsilon_F)$ \cite{our-rel}, 
therefore the integral in Eq.\ (\ref{t2f}) has a logarithmic divergence. 
At the upper limit, this integral is cut at $\epsilon-\epsilon_F\sim\Delta\mu$, since this is 
the maximum energy provided by the source. At the lower limit, the integral has to be cut at 
$\epsilon-\epsilon_F\sim T\Delta\mu$, due to broadening of the distribution function induced by the noise.    
This leads to the  behavior $\Theta\propto  T\ln (1/T)\Delta\mu$ at $T\ll 1$, i.e., the noise temperature 
is singular in $T$ and linear in $\Delta\mu$, contrary to the prediction of Ref.\ \cite{neder-wrong}. 
In the rest of the paper, we demonstrate this fact rigorously by
resumming weak tunneling using the non-equilibrium bosonization technique \cite{our-phas},
and investigate various physical regimes in detail.    

{\it Model and theoretical method.}--- In an experiment, the applied voltage bias $\Delta\mu$ is typically
much smaller than the Fermi energy $\epsilon_F$. Thus, it is appropriate to use the low-energy effective theory 
\cite{eff-theory} describing edge states at filling factor $\nu =2$ as collective fluctuations of the charge
density $\rho_{s\alpha}(x)$, where $\alpha= 1,2$ enumerates channels at the QH edge, 
and  $s = L,M,U$ denotes the lower, middle, and upper edge (see Fig.~\ref{tun}).
The charge density fields are expressed in terms of chiral boson fields, $\phi_{s\alpha}(x)$,
satisfying the commutation relations
\begin{equation}
[\phi_{s\alpha}(x),\phi_{r\beta}(y)]=i\pi\delta_{sr}\delta_{\alpha\beta}{\rm sgn}(x-y),
\label{fields}
\end{equation}
namely, $\rho_{s\alpha}(x)=(1/2\pi)\partial_x\phi_{s\alpha}(x)$.  
The total Hamiltonian of the system, $\mathcal{H}=\mathcal{H}_0+(A+A'+{\rm h.c.})$, 
contains the term describing the edge states 
\begin{equation}
\hspace*{-1pt}\mathcal{H}_0 = \frac{1}{8\pi^2}\!\sum_{s,\alpha,\beta} \!\int\! dxdy V_{\alpha\beta}(x-y)
\partial_x\phi_{s\alpha}(x)\partial_y\phi_{s\beta}(y),
\label{hamilt}
\end{equation}
where the kernel, $V_{\alpha\beta}(x-y) = 2\pi v_F \delta_{\alpha\beta}\delta(x-y) + U_{\alpha\beta}(x-y)$,
includes the free-fermion contribution with the Fermi velocity $v_F$, and the Coulomb interaction potential $U_{\alpha\beta}$.
Vertex operators
\begin{equation}
A= t\, e^{i\phi_{L1}(0)-i\phi_{M1}(0)},  \quad A'= t' e^{i\phi_{M2}(D)-i\phi_{U2}(D)}
\end{equation}
describe electron tunneling between the edge channels at the QPCs. The right QPC, serving as a non-invasive 
detector, is in the weak tunneling regime. Therefore, we treat corresponding operator $A'$ perturbatively
\cite{foot-exact}.  

The backscattering current at the second QPC may be written as $j_{\rm bs}=i(A'-A'^{\dagger})$
and, to the leading order in the tunneling operator $A'$,
the noise power (\ref{zfnp}) of this current reads: 
$S_{\rm bs} = \int dt \langle\{A'^\dagger(t),A'(0)\} \rangle$. The relatively straightforward
steps lead to the standard result (\ref{useful}), and to the effective 
noise temperature (\ref{t2f}) with
\begin{subequations}
\begin{eqnarray}
\label{temp}
f(\epsilon) &\propto& \int dt e^{-i(\epsilon-\epsilon_F) t}K(t),\\
K(t) &=& \langle e^{-i\phi_{M2}(D,t)}e^{i\phi_{M2}(D,0)}\rangle,
\label{corrdef} 
\end{eqnarray}
\end{subequations}
where the normalization prefactor in (\ref{temp}) is determined by 
the condition $f(\epsilon)=1$ at $\epsilon\to-\infty$. 
The average in the definition of $K(t)$ has to be taken with respect to the
\textit{non-equilibrium} state created by the source QPC. Therefore, we apply the 
non-equilibrium bosonization technique proposed in our earlier work \cite{our-phas}. 

The Hamiltonian (\ref{hamilt}), together with the commutation relations (\ref{fields}),
generates equations of motion for the fields $\phi_{s\alpha}$ that have to be accompanied with 
boundary conditions:
\begin{subequations}
\label{eom}
\begin{eqnarray}
\hspace*{-0.22cm}\!\partial_t\phi_{M\alpha}(x,t)\! =\! -\frac{1}{2\pi}\!\sum_\beta\!\int\!\! dy V_{\alpha\beta}(x-y)\partial_y\phi_{M\beta}(y,t),
\label{eoma}\\
\partial_t\phi_{M\alpha}(0,t)\! =\!- 2\pi j_{\alpha}(t).
\label{eomb}
\end{eqnarray}
\end{subequations}
We place the boundary at the point $x=0$, right after the source QPC.
At low energies of interest, the characteristic length scales are much longer
than the screening length $d$ of the Coulomb 
potential $U_{\alpha\beta}(x-y)$. Therefore, we can neglect its logarithmic dispersion and approximate
$U_{\alpha\beta}(x-y) = U_{\alpha\beta}\delta (x-y)$, 
and consequently, $V_{\alpha\beta}(x-y) = V_{\alpha\beta}\delta (x-y)$. 
Then, Eqs.\ (\ref{eom}) acquire a form of first-order differential equations. 
We solve these equations  
by diagonalizing the matrix $\hat{V}\equiv V_{\alpha\beta}$ with the rotation  
$\hat{V} = \hat{S}(\theta)\hat\Lambda \hat{S}^\dag(\theta)$ by the angle $\theta$ defined as 
$\tan\! 2\theta=2V_{12}/(V_{11}-V_{22})$. Then, the spectrum of the collective charge excitations splits in two modes, $\hat\Lambda = {\rm diag}(u,v)$, with the speeds
$u,v = (V_{11}+V_{22})/2\pm\sqrt{(V_{11}-V_{22})^2/4+V_{12}^2}$. 
Imposing the boundary condition (\ref{eomb}), we arrive at the solution 
\begin{subequations}
\begin{multline}
\label{phis}
\phi_{M2}(x,t)=  \lambda_1Q_{1}(t_u)+\lambda_2Q_{2}(t_u)\\-\lambda_1Q_{1}(t_v)+\lambda'_2Q_{2}(t_v), 
\end{multline}\vspace{-7mm}
\begin{equation}
 \lambda_1 = \pi\sin 2\theta,\; \lambda_2 = \pi(1+\cos 2\theta),\; \lambda'_2 = 2\pi-\lambda_2,
\label{lam}
\end{equation}
\label{twoeqs}
\end{subequations}
where we have introduced the injected charges $Q_{\alpha}(t)=\int_{-\infty}^t dt' j_{\alpha}(t')$, and 
notations $t_u = t- x/u$ and  $t_v = t- x/v$. 

Since the edge state dynamics is chiral, and the screened Coulomb interaction is effectively short-range,
the fields $\phi_{M\alpha}$ 
do not influence fluctuations of the currents $j_{\alpha}$ at the QPC \cite{Sukh-Che,our}. 
As a consequence, the electron transport through a single QPC is not affected by the interaction,
which seems to be an experimental fact \cite{Basel}. Therefore, when finding the correlator 
(\ref{corrdef}), one may utilize the free-fermion scattering theory for the statistics
of injected charges  $Q_{\alpha}$
\cite{Blanter,Levitov}.

{\it Gaussian noise regime.}--- 
It has been shown in Ref.\ \cite{our-rel} that a weak dispersion
of plasmon modes suppresses higher-order cumulants at large distances.
Therefore, we first focus on the situation, where the fluctuations
of the boson fields may be considered Gaussian. Then, the logarithm of the correlation function
(\ref{corrdef}) can be written as
$\ln K(t) = - \langle
\phi_{M2}^2(D,t) -
2\phi_{M2}(D,t)\phi_{M2}(D,0) + \phi_{M2}^2(D,0)\rangle/2$,
where  the linear terms in field $\phi_{M2}$ vanish, since there is no contribution
to the average current in the inner channel.
Using Eqs.\ (\ref{twoeqs}), we obtain:
\begin{multline}
\ln K(t) = -2\pi \!\int \frac{d\omega}{\omega^2}(1-e^{i\omega
t})\Big\{\Big(\frac{\lambda_1}{\pi}\Big)^2\!\sin^2\!\Big(\frac{\omega t_D}{2}\Big)S_1(\omega) \\
+\Big[1-\frac{\lambda_2\lambda'_2}{\pi^2}\sin^2\Big(\frac{\omega t_D}{2}\Big)\Big]S_2(\omega)\Big\},
\label{t1}
\end{multline}
where we have introduced the noise power, $S_\alpha(\omega) =
\int dt e^{i\omega t}\langle\delta j_\alpha(t)\delta
j_\alpha(0)\rangle$, and the time delay between the wave packets, $t_D = D/v-D/u$. 

Since the transport through the injecting QPC is not affected by interactions, 
the free-fermion scattering approach \cite{Blanter} may be used to obtain
\begin{equation}
S_\alpha(\omega) = S_{\rm q}(\omega) + T_\alpha(1-T_\alpha)  S_{\rm n}(\omega),
\end{equation}
where $S_{\rm q}(\omega) = \omega\theta(\omega)/2\pi$ is the ground-state
(Fermi sea) contribution, and
$S_{\rm n}(\omega) = \sum_\pm [S_{\rm q}(\omega\pm\Delta\mu)-S_{\rm q}(\omega)]$
is the  non-equilibrium part.
Therefore, in the  expression (\ref{t1}) the ground-state and non-equilibrium contributions
separate, 
$\ln K(t) = -\ln \epsilon_F t+\ln K_{\rm n}(t)$,  and 
the noise temperature (\ref{t2f}) may be presented as
\begin{equation}
\Theta=-\frac{1}{2\pi\ln 2}\int\frac{dt}{(t-i\eta)^2}\,K_{\rm n}(t),\quad\eta\to 0,
\label{t2f-2}
\end{equation}
where the non-equilibrium contribution reads
\begin{multline}
\ln K_{\rm n}(t) = -4 T(1-T)(\lambda_1/\pi)^2\\
\times\int_0^1 \frac{dx}{x^2}(1-x)\sin^2\!
\Big(\frac{\Delta\mu t x}{2}\Big)\sin^2\!\Big(\frac{\Delta\mu t_D x}{2}\Big).
\label{second}
\end{multline}
We note, that the ground-state contribution to the correlator $K$
is always Gaussian and is independent of the interactions, because
the effect of the injecting QPC on the states below $\epsilon=\epsilon_F$ is simply a
unitary transformation. 

Next, we focus on the weak injection regime, $T\ll 1$, verify the validity of the perturbation 
approach with respect to weak tunneling, and show that it may fail. 
It turns out, that the expansion of $K_{\rm n}$ with
respect to $T$ as $K_{\rm n}=1+\ln K_{\rm n}+\ldots$ is dangerous, because $\ln K_{\rm n}$
diverges at large $t$ and $t_D$. More precisely,
at distances $D\gg D_{\rm ex}$, where $D_{\rm ex}\equiv uv/[(u-v)\Delta\mu]$ is the characteristic 
length of the energy exchange between edge channels \cite{our-rel},
its asymptotic reads: $\ln K_{\rm n}=
 -(\lambda_1^2/2\pi)T\Delta\mu\min(t,t_D)$.
Therefore, to leading order in tunneling at the first QPC, the time integral in Eq.\ (\ref{t2f-2})
diverges logarithmically. At the short-time limit, this integral should be cut at $t\sim 1/\Delta\mu$,
where it behaves regularly.
At the upper limit, it is cut at either $t\sim 1/(T\Delta\mu)$, where $\ln K_{\rm n}$ is
not small, and perturbation approach fails, or at $t\sim t_D$, where $\ln K_{\rm n}$ 
takes a constant value smaller than $1$ if $T\Delta\mu t_D\ll 1$. Thus, 
for $T\ll 1$ the noise temperature reads
\begin{equation}
\frac{\Theta}{\Delta\mu} = \frac{\lambda_1^2T}{2\pi^2\ln 2}\left\{
\begin{array}{ll}
\ln(\Delta\mu t_D), & \mbox{if $D_{\rm ex}/T\gg D\gg D_{\rm ex}$},\\
\vspace*{-0.2cm}\\
\ln(1/T), & \mbox{if $D\gg D_{\rm ex}/T$}.
\end{array}
\right.
\label{t2f-3}
\end{equation}
We recall the notations $t_D=D/v-D/u$ and $D_{\rm ex}= uv/[(u-v)\Delta\mu]$.

It remains to investigate the noise temperature at  short distances, $D\ll D_{\rm ex}$.
In this case, we can replace $\sin^2(\Delta\mu t_D x/2)\to (\Delta\mu t_D x/2)^2$ in 
Eq.\ (\ref{second}). It is more convenient to substitute $\ln K_{\rm n}$ into 
Eq.\ (\ref{t2f-2}) and first evaluate the time integral, and then the integral over $x$.
The result for the noise temperature reads:
\begin{equation}
\Theta=\frac{\lambda_1^2Tt_D^2}{24\pi^2\ln2}(\Delta\mu)^3,\quad D\ll D_{\rm ex}.
\label{t2f-4}
\end{equation}
This regime can be viewed as perturbative both with respect to tunneling and
 interactions.

{\it Non-Gaussian noise: exact results.}--- 
To complete our analysis, we investigate the situation, 
where even at long distances, $D\gg D_{\rm ex}/T$, the fluctuations
of the edge fields remain non-Gaussian. At such distances,
two plasmon modes, arriving with the time delay $t_D$ 
longer than the correlation time $1/\Delta\mu$ of boundary currents (see Fig.\ \ref{conc}),
separate the injected charges $Q_\alpha$ in Eq.\ (\ref{phis}) into uncorrelated terms. 
Therefore, the correlation function $K$ splits in the product of four terms  
\begin{equation}
K(t) = \chi_{1}(\lambda_1,t)\chi_{1}(-\lambda_1,t)\chi_{2}(\lambda_2,t)\chi_2(\lambda'_2,t),
\label{cf}
\end{equation}
each taking the form of the generator of full counting statistics (FCS) \cite{Levitov}: 
\begin{equation}
\chi_{\alpha}(\lambda,t)=\langle e^{i\lambda Q_{\alpha}(t)}e^{-i\lambda Q_{\alpha}(0)}\rangle.
\label{fcs}
\end{equation}
The correlation function (\ref{cf}) is independent of $D$, i.e., in 
the limit $D\gg D_{\rm ex}/T$ electrons in the inner channel do indeed reach a non-trivial
stationary state.

We note, that the FCS generator of the inner channel at the edge $M$  contains only 
the Gaussian contribution from the Fermi sea,
$\ln\chi_2(\lambda, t)=-(\lambda^2/4\pi^2)\ln \epsilon_Ft$, while 
the FCS generator at the outer channel, being perturbed by a QPC, acquires 
additional non-Gaussian part from the transport electrons, 
$
\ln\chi_1(\lambda, t) = -(\lambda^2/4\pi^2)\ln \epsilon_Ft +\ln\chi_{\rm n}(\lambda, t).  
$
This leads to the expression (\ref{t2f-2}) for the effective noise temperature 
with  
\begin{equation}
K_{\rm n}(t) = \chi_{\rm n}(\lambda_1,t)\chi_{\rm n}(-\lambda_1, t). 
\label{temp2}
\end{equation}
We stress that in the limit $\Delta\mu\ll\epsilon_F$ the non-equilibrium FCS generator $\chi_{\rm n}$
depends on time only via the dimensionless combination $\Delta\mu t$, which is the consequence 
of a free-fermion character of the electron transport through a single QPC. 
Therefore, at distances $D\gg D_{\rm ex}/T$ the noise temperature is always linear in applied voltage 
bias $\Delta\mu$, independently of details of the interaction. 

\begin{figure}[h]
\begin{center}
\epsfxsize=8cm
\epsfbox{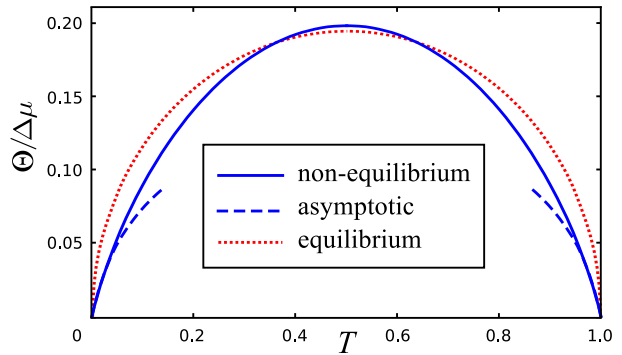}
\end{center}
\vspace{-3mm}
\caption{(Color on-line) The normalized effective noise temperature at the detector QPC
is plotted as a function of the transparency $T$ of the source QPC, generating a non-Gaussian noise. 
{\it Solid, blue line}: The exact value of $\Theta/\Delta\mu$ for a non-equilibrium
stationary state at $D\gg D_{\rm ex}/T$, evaluated with the help of the determinant representation of the FCS 
generator \cite{Levitov}.  
{\it Dashed, blue line}: Its asymptotic behavior (\ref{f-res}) for $T\ll 1$, and similar
result for $1-T\ll 1$.
{\it Dotted, red line}: 
The temperature (\ref{eq-t}) of a locally equilibrium state at $D\to\infty$, extracted from the 
energy flux in the inner edge channel.} \vspace{-3mm}
\label{noise-n}
\end{figure}

In the following, we concentrate on the realistic case of a Coulomb interaction
screened at distances $d\gg a$, where $a$ is the distance between edge channels.
Therefore, one may approximate $U_{\alpha\beta} =\pi u$, where $u/v_F\sim\ln(d/a)\gg 1$, 
so that $\theta = \pi/4$ and $\lambda_1 = \pi$ \cite{our}. 
The dimensionless function $\chi_{\rm n}(\pi, t)$ 
can be represented as a determinant of a single-particle operator \cite{Levitov} and calculated numerically \cite{our-num}.
The result for the normalized noise temperature $\Theta/\Delta\mu$ as a function of transparency 
$T$ of the injecting QPC is shown in Fig.\ \ref{noise-n}. We also plot the normalized temperature 
$\Theta_{\rm eq}/\Delta\mu$ of an equilibrium distribution reached by electrons in the inner channel
at $D\to\infty$,
\begin{equation}
 \Theta_{\rm eq}/\Delta\mu = \sqrt{3T(1-T)/2\pi^2},
\label{eq-t}
\end{equation}
which is found by comparing the energy flux of electrons $\pi^2\Theta_{\rm eq}^2/6$ 
to the half of the heat flux $\Delta\mu^2 T(1-T)/2$ injected at the first QPC.

One can see in Fig.\ \ref{noise-n} a singular behavior of $\Theta$ at $T\to 0$
and $T\to 1$. In order to describe it analytically, we recall the FCS generator for 
the tunneling process: 
$
\ln\chi_{\rm n}(\lambda_1, t) = (\Delta\mu |t|/2\pi)T(e^{i\lambda_1}-1)$
for $ \Delta\mu t \gg 1$. 
Note, that this FCS generator is universal, i.e., it does not require an assumption of 
free-electron transport at the QPC, and reflects the simple fact that 
tunneling is a Poisson process with all the current cumulants 
equal to the average current.
Substituting this expression into Eqs.\ (\ref{t2f-2}) and (\ref{temp2}), 
and setting $\lambda_1=\pi$, we find the noise temperature
at $T\ll 1$ in the non-Gaussian noise regime
\begin{equation}
\Theta/\Delta\mu = (2/\pi^2\!\ln 2)\, T\ln(1/T),\quad D\gg D_{\rm ex}/T.
\label{f-res}
\end{equation}
It differs from the one for the Gaussian noise, Eq.\ (\ref{t2f-3}), only by a numerical 
prefactor.

To summarize, we have investigated the effects of the integrability of the charge dynamics at 
QH edge at filling factor $\nu=2$, where two chiral edge channels coexist. We have found that 
the double-step electron distribution, created in one of the channels with the help of a voltage-biased 
QPC, evolves via several intermediate regimes to a non-equilibrium stationary state.  
Measuring the backscattering current noise in the second, co-propagating channel
reveals a non-trivial effect of the integrability and strong inter-channel Coulomb interactions:  
The effective noise temperature $\Theta$ of this stationary state is a non-analytical function of 
the transparency $T$, which scales as $\Theta \propto T\ln(1/T)$ at $T\ll 1$. 

We acknowledge support from the Swiss NSF.

\bibliographystyle{apsrev}

\begin{thebibliography}{99}
\bibitem{foot-exact} For arbitrary $T'$, the exact calculations  
show that the noise temperature is equal to $1-T'$ times its perturbative 
value. Therefore, to avoid unnecessary complications, we use the perturbative 
approach.
\end{thebibliography}

\end{document}